\documentclass[twocolumn,showpacs,preprintnumbers,prb,amsmath,amssymb]{revtex4}

\usepackage{graphicx}% Include figure files
\usepackage{dcolumn}% Align table columns on decimal point
\usepackage{bm}% bold math
\usepackage{epsfig}

\begin{document}

\title{
Andreev magnetotransport in low-dimensional proximity structures:\\
Spin-dependent conductance enhancement}
\author{Grygoriy Tkachov and Klaus Richter}

\affiliation{Institute for Theoretical Physics, Regensburg University, 
93040 Regensburg, Germany}

\date{\today}

\begin{abstract} 
We study the excess conductance due to the superconducting proximity effect 
in a ballistic two-dimensional electron system 
subject to an in-plane magnetic field.
We show that under certain conditions the interplay of 
the Zeeman spin splitting and the effect of a screening supercurrent 
gives rise to a spin-selective Andreev enhancement of the conductance 
and anomalies in its voltage, temperature and magnetic field characteristics. 
The magnetic-field influence on Andreev reflection is discussed 
in the context of using superconducting hybrid junctions for spin detection.
\end{abstract}

\pacs{74.45.+c, 73.23.Ad, 72.25.Dc}

\maketitle

Advances in nanotechnology of semiconductor-superconductor 
junctions~\cite{Nano} have created a unique opportunity to investigate the interplay 
between superconducting phase-coherence 
and various electronic properties of low-dimensional semiconductors. 
Recent developments in this field include 
the realization of a long-range Josephson coupling mediated 
by Andreev reflection~\cite{Andreev} of ballistic two-dimensional 
electrons and controlled by the injection of hot carriers~\cite{Jo,Hot}, 
the observation of a giant proximity-induced enhancement 
of the conductance of two-dimensional electron 
systems (2DES)~\cite{Giant}, 
theoretical~\cite{Billiards} and experimental~\cite{Eroms} studies 
of Andreev billiards, classical~\cite{Uhlisch} and quantum~\cite{Edges,Eroms1} Andreev edge 
states in high magnetic fields. 

In this paper we discuss a possibility of using the superconducting 
proximity effect for detecting the spin of transport carriers in low-dimensional systems, 
a problem closely related to the ongoing work on spin injection in 
semiconductors~\cite{Spin}.  
In nonmagnetic normal metal-superconductor (NS) junctions
spin resolving transport measurements were first reported in Ref.~\onlinecite{MTF} 
where the magnetic field spin splitting of the quasiparticle density of 
states in thin superconducting films served as an electron filter. 
Applied to ferromagnet-superconductor systems, this idea has developed into 
a sensitive technique of analyzing the spin polarization of 
ferromagnetic metals~\cite{TM}.
We note that the findings of Ref.~\onlinecite{MTF} 
are specific to low-transparency tunnel junctions where the superconducting proximity effect 
is negligible and hence the electron transport is predominantly a quasiparticle one.

\begin{figure}[!t]
\begin{center}
\epsfxsize=0.6\hsize
\epsffile{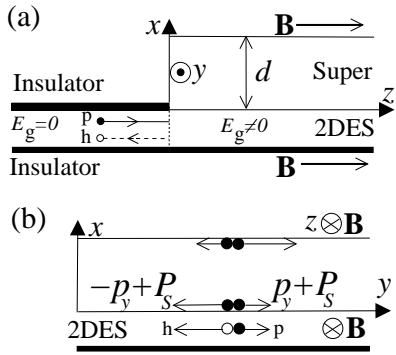}
\end{center}
\caption{
(a) Cross section of a planar superconductor-2DES contact.
At energies below the minigap $E_g$ 
a particle (p) incident from the normal part of the 2DES is 
reflected as a hole (h) giving rise to conductance enhancement.  
(b) Schematic of the supercurrent flow in the $xy$ plane. 
The particle and hole momenta in the 2DES are both 
shifted by $P_S=-(e/c)Bd/2$ in order to match the Cooper pair momentum 
$2P_S$. 
}
\label{Geo}
\end{figure}

In structures with improved interfacial quality the penetration of the superconducting order 
parameter into the normal system is accompanied by the conversion of a quasiparticle current 
into a supercurrent via Andreev reflection which manifests itself as a low-bias conductance
enhancement~\cite{Cast,vanWees,Petrashov,Green,Volkov,Pannetier,Gap,Been,Imry}.   
One of the most striking examples of the proximity effect 
occurs in ballistic semiconductor quantum wells (2DES) with a lateral superconducting contact 
[see Fig.~\ref{Geo}(a)].
In this case the conductance in the plane of the quantum well 
can nearly twice exceed the normal-state value~\cite{Giant,Eroms,Eroms1} 
due to a proximity-induced mixing of particle and hole states 
characterized by a superconducting minigap $E_g$ in the excitation spectrum 
of the 2DES~\cite{Volkov}. 
For such proximity structures 
we study the spin dependence of Andreev reflection arising from the Zeeman splitting of the gapped 
states in the 2DES subject to an in-plane magnetic field ${\bf B}$.

Normally, a noticeable spin splitting in superconductors 
requires rather strong magnetic fields (above $1$T)~\cite{MTF} 
of the order of the field $B_c\approx\Delta/\mu_B$ 
corresponding to the paramagnetic limit~\cite{Para} ($\Delta$ and $\mu_B$ are 
the gap energy in the superconductor and the Bohr magneton).
The advantage of the low-dimensional proximity structures is that 
the scale of relevant fields is set by the minigap, $B_g\approx (2E_g/g\Delta)B_c$, 
and hence can be much smaller than $B_c$ both due to $E_g\ll\Delta$ 
and large electron $g$-factors (e.g. $|g|\approx 10-14$ in InAs/AlSb quantum wells~\cite{Nano,g}).

Apart from the smaller splitting fields, the magnetic field influence on 
ballistic proximity systems has one more specific feature. 
Because of the screening supercurrent generated by the field, 
the quasiparticles in the 2DES can acquire a 
significant Galilean energy shift arising from a finite Cooper pair 
momentum at the superconductor surface~\cite{Thermo}. 
In ballistic quantum wells with weak momentum scattering
the Galilean supercurrent effect does not average out, 
unlike in conventional diffusive superconductors~\cite{MTF,Maki,Anthore}, 
and therefore must be taken into account along with the Zeeman splitting.
Here we discuss the influence of these two competing magnetic field effects 
on the voltage, temperature and magnetic-field characteristics of Andreev 
transport.

We consider a 2DES coupled to a superconducting film via a barrier 
of low transparency ${\cal T}\ll 1$ [Fig.~\ref{Geo}(a)]. 
The film thickness $d$ is assumed much smaller 
than both the superconducting coherence length and the London 
penetration depth in a parallel field ${\bf B}=[0; 0; B]$. 
In a ballistic quantum well the motion of particles and holes 
with energies smaller than $\Delta$ is coupled in the proximity region 
[$z>0$ in Fig.~\ref{Geo}(a)] via multiple Andreev reflections.
In the quantum limit (2DES) the 
correlated particle-hole motion can be described 
in terms of the effective pairing energy which gives 
rise to the quasiparticle minigap $E_g\approx (v_F/v_{F_S}){\cal T}E_0\ll\Delta$~\cite{Volkov}. 
It depends on the energy $E_0$
of the lowest occupied subband, the transparency 
${\cal T}$ and the ratio of the Fermi velocities in the normal ($v_F$) and 
superconducting ($v_{F_S}$) systems. 
Accordingly, electron scattering from the 
proximity region can be treated
using effective Bogolubov-de Gennes equations for the electron 
$u_{\alpha}(z)$ and hole $v_{-\alpha}(z)$ 
wave functions averaged over the thickness of the 2DES~\cite{BdG}:
\!
\begin{eqnarray}
\left[
\begin{array}{cc}
\tilde\epsilon+\frac{\hbar^2\partial_z^2+\tilde p_F^2}{2m} & 
-E_g\Theta(z)\\
E_g\Theta(z) &
-\tilde\epsilon+\frac{\hbar^2\partial_z^2+\tilde p_F^2}{2m}
\end{array}
\right]
\left[
\begin{array}{cc}
u_{\alpha}(z) \\
v_{-\alpha}(z) 
\end{array}
\right]=0.
\label{BdG}
\end{eqnarray}
Here $\partial_z$ stands for a derivative; $m$, $p_F$, and $\alpha=\pm 
1/2$ are respectively the electron mass, Fermi momentum, and spin;
$\tilde p_F=(p_F^2-p_y^2)^{1/2}$ where
the parallel momentum $p_y=p_F\sin\theta$ depends on the angle $|\theta|\leq \pi/2$ 
measured from the $z$ axis in the 2DES plane. 
As $E_g$ scales with the transparency ${\cal T}$,
there is virtually no "leaking" of electron pairing into  
the "normal" region [$z<0$ in Fig.~\ref{Geo}(a)] because it would 
require Cooper pair tunneling through the thick insulator. Hence we 
assume the step-like pairing energy $E_g\Theta(z)$ with $\Theta(z)$ being the 
Heaviside function. 

A weak magnetic field can be taken into account by an energy shift:
\!
\begin{equation}
\tilde\epsilon =\epsilon +\alpha g\mu_B B -v_FP_S\sin\theta,\quad 
P_S=-(e/c)Bd/2,	
\label{shift}
\end{equation}
where the second term is the Zeeman energy whereas  
the third one is a Galilean energy arising from the shift 
of both particle and hole momenta which accounts for   
a screening supercurrent generated in the superconductor by 
a parallel magnetic field [see Fig.~\ref{Geo}(b)].
As the thickness of the 2DES is considered negligible
compared to $d$, the momentum shift of electrons and holes 
in the 2DES can be taken equal to the surface Cooper pair
momentum per electron, $P_S$, in Eq.~(\ref{shift}) ($e>0$ is the 
elementary charge). 
It is proportional to half of the film thickness 
reflecting the fact that the field fully penetrates the film 
and generates an antisymmetric (linear) distribution of the 
supercurrent density with respect to its middle plane.

\begin{figure}[!t]
\begin{center}
\epsfxsize=0.65\hsize
\epsffile{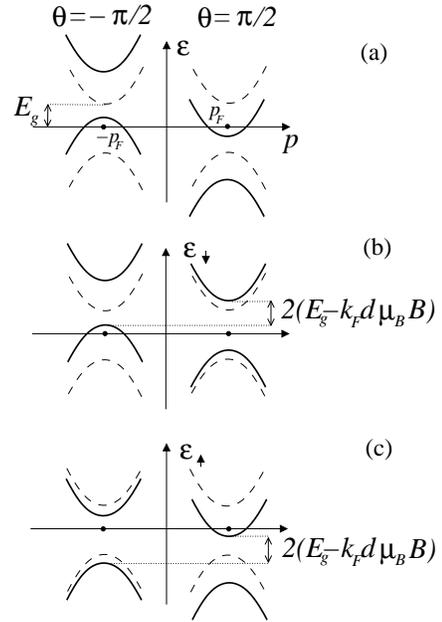}
\end{center}
\caption{
Schematic view of the excitation spectrum (\ref{Spectr})
in the direction ($\theta=\pi/2$) and opposite
($\theta=-\pi/2$) to the supercurrent at $B>B_g$:
(a) - for spinless quasiparticles ($\tilde g=0$), (b) and (c) - for 
spin-down and spin-up ones with $\tilde g>1$. 
Dashed curves correspond to $B=0$. 
}
\label{E}
\end{figure}

For $|\tilde\epsilon|\ll E_F=p^2_F/2m$ the magnetic field effect on the 
quasiparticle energies is important only in the proximity region.
Indeed, from Eqs.~(\ref{BdG}) and~(\ref{shift}) one finds the 
excitation spectrum in this region as
\!
\begin{eqnarray}
\epsilon_{\alpha p\theta}^\pm=-(\alpha g+k_Fd\sin\theta)\mu_B B 
\pm[v_F^2(p-p_F)^2+E_g^2]^{\frac{1}{2}},
\label{Spectr}
\end{eqnarray}
where $p$ is the absolute value of the momentum, and $k_F=p_F/\hbar$.
The magnetic field shifts the energies~(\ref{Spectr}) 
with respect to the Fermi level so that 
at a certain field  
\!
\begin{eqnarray}
B_g=E_g/[\mu_B k_Fd (\tilde g +1)],\qquad \tilde g=g/2k_Fd, 
\label{Bg}	
\end{eqnarray}
the excitations with momenta (anti)parallel to the supercurrent ($\theta =\pm\pi/2$) become 
gapless. We point out that in Eqs.~(\ref{Spectr}) and ~(\ref{Bg}) 
the interplay of the Zeeman and supercurrent effects is controlled by a single material- and geometry-dependent parameter
$\tilde g$. This is shown in Fig.~\ref{E}
where we sketch the $p$-dependence 
of the energies $\epsilon_{\alpha p\theta}^\pm$ near the Fermi 
points $\pm p_F$ in the direction ($\theta =\pi/2$) and 
opposite ($\theta =-\pi/2$) to the supercurrent. 
For $\tilde g\ll 1$ (thick superconductors or small $g$-factors)
the quasiparticles can be treated as spinless, 
and the minigap vanishes as soon as the 
dispersion curves cross the Fermi energy simultaneously at $\pm p_F$ 
[Fig.~\ref{E}(a)]. For $\tilde g>1$ (thin superconductors and/or large 
$g$-factors) the Zeeman splitting preserves the minigap 
at finite energies even after the appearance of gapless 
excitations at the Fermi level [Figs.~\ref{E}(b) and (c)]. 
The gap remains open in the field range $1<B/B_g<1+\tilde g$, 
rather broad for $\tilde g>1$.

\begin{figure}[t]
\begin{center}
\epsfxsize=0.8\hsize
\epsffile{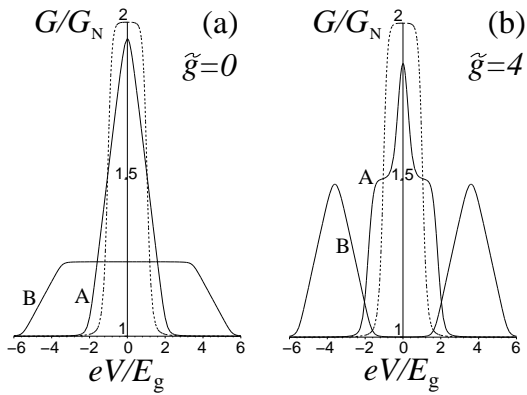}
\end{center}
\caption{
Conductance versus voltage for (a) spinless and (b) spin-polarized 
electrons at magnetic fields $B/B_g$: 
$0$ (dashed), $1$ (A) and  $4.5$ (B); $k_BT/E_g=0.1$.}
\label{G_V}
\end{figure}

Due to the superconducting proximity effect
the current of quasiparticles entering from 
the normal region [$z<0$ in Fig.~\ref{Geo}(a)]
can be converted into a supercurrent via Andreev reflection 
at low energies $|\tilde\epsilon|<E_g$.
The efficiency of the Andreev conversion is insured by  
the lack of a barrier between the 
normal and proximity-affected regions of the 2DES.
That is why for calculating the conductance we can use the solution of the 
scattering problem
of Ref.~\onlinecite{BTK} applied to an ideal NS interface.
In reality, normal reflection does occur 
because the contact to a metal slightly modifies 
the quantum well resulting in a mismatch of the 
Fermi momenta~\cite{Eroms} at $z=0$. 
However, for ${\cal T}\ll 1$ this mismatch is rather weak. 
We have checked that it has a small effect 
on the zero-field conductance 
while we have found no significant difference at finite fields $B\sim B_g$, the regime we 
are interested in.

In the absence of normal scattering the probability 
$A(\tilde\epsilon)$  
for a particle to be reflected from the proximity region as a hole 
is given within the usual approximation $|\tilde\epsilon|, E_g\ll 
E_F\cos^2\theta$ by~\cite{BTK}
\! 
\begin{eqnarray}
A(\tilde\epsilon)=
\Theta\left(E_g-|\tilde\epsilon|\right)
+
\Theta\left(|\tilde\epsilon|-E_g\right)
\frac{|\tilde\epsilon|-[\tilde\epsilon^2-E_g^2]^{\frac{1}{2}}}
{|\tilde\epsilon|+[\tilde\epsilon^2-E_g^2]^{\frac{1}{2}}}.
\label{A}
\end{eqnarray}
The expression for the conductance reads~\cite{Approx}
\!
\begin{eqnarray}
&&
\frac{G(V,B,T)}{G_N}=1+\frac{1}{4}\sum\limits_\alpha
\int d\epsilon\left(-\frac{\partial f(\epsilon 
-eV)}{\partial\epsilon}\right)
\times
\nonumber\\
&&
\times
\int_{-\pi/2}^{\pi/2}
d\theta\cos\theta 
A\left[\epsilon +(\alpha g+k_Fd\sin\theta)\mu_B B \right],
\label{G}
\end{eqnarray}
where $G_N$ is the conductance of the normal 2DES and 
$f(\epsilon -eV)$ is the Fermi distribution of the incident electrons at 
bias energy $eV$. 
Since the $\theta$-dependence enters through the energy shift in $A$, the double 
integral in Eq.~(\ref{G}) can be easily reduced to a single one:
\!
\begin{eqnarray}
&&
\frac{G(V,B,T)}{G_N}=1+\sum\limits_\alpha
\int_{-\infty}^\infty 
\frac{d\tilde\epsilon\, A(\tilde\epsilon)}{4k_Fd\mu_B B} 
\times
\label{G1}\\
&&
\times
\left[
f\left(\tilde\epsilon -eV - (\alpha g+k_Fd)\mu_B B \right)-
\right.
\nonumber\\
&&
\left.
-
f\left(\tilde\epsilon -eV - (\alpha g-k_Fd)\mu_B B \right)
\right].
\nonumber
\end{eqnarray}

\begin{figure}[t]
\begin{center}
\epsfxsize=0.8\hsize
\epsffile{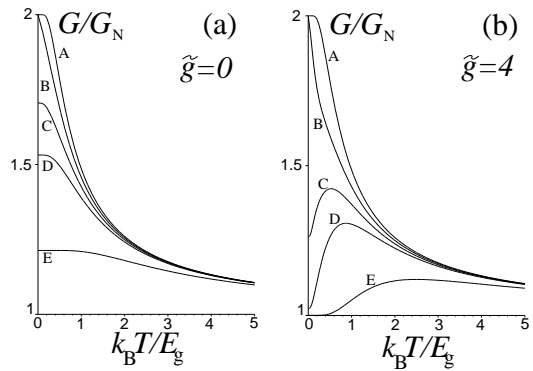}
\end{center}
\caption{
Zero-bias conductance versus temperature for (a) spinless and (b) 
spin-polarized electrons at magnetic fields $B/B_g$: 
$0$ (A), $1$ (B), $1.5$ (C), $2$ (D), and $5$ (E).}
\label{G_T}
\end{figure}

Figure~\ref{G_V}(a) shows the suppression of 
the excess conductance peak due to the vanishing of the minigap for 
spinless electrons 
as the field exceeds $B_g$. In contrast to this, for spin-polarized 
electrons   
the zero-bias anomaly splits into two twice smaller peaks 
[Fig.~\ref{G_V}(b)]
because the spin-dependent minigap in the excitation spectrum, 
Eq.~(\ref{Spectr}), still supports 
Andreev reflection of spin-down particles for $V>0$
and spin-up holes for $V<0$. In other words, for a given voltage
only quasiparticles of one spin-orientation give rise to the 
conductance enhancement. The spin-dependent Andreev process
persists until the minigap closes at $B=B_g(1+\tilde g)$ followed
by an overall decrease in the excess conductance at higher fields.

\begin{figure}[b]
\begin{center}
\epsfxsize=0.75\hsize
\epsffile{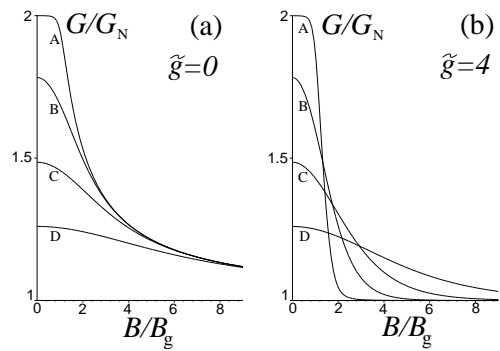}
\end{center}
\caption{
Zero-bias magnetoconductance for (a) spinless and (b) spin-polarized 
electrons at temperatures $T/(k^{-1}_BE_g)$: 
$0.1$ (A), $0.5$ (B), $1$ (C), and  $2$ (D).}
\label{G_B}
\end{figure}

The temperature dependence of the zero-bias conductance, shown in
Fig.~\ref{G_T}(b), reveals one more manifestation of the magnetic spin splitting: 
a maximum in $G(T)$ in contrast to a monotonic decrease for spinless electrons 
[Fig.~\ref{G_T}(a)].
It is due to the shift of the minigap to finite energies where Andreev reflection 
is mediated by thermally excited quasiparticles. 
The corresponding behaviour of the zero-bias magnetoconductance at 
different temperatures 
is shown in Figs.~\ref{G_B}(a) and (b). The energy and $B$-field 
anomalies discussed above are already well-resolved when 
$\tilde g$ exceeds~1.

To conclude we have demonstrated that the magnetic field spin splitting 
of the quasiparticle states in proximity S-2DES nanostructures
results in a pronounced spin-dependent Andreev 
reflection and anomalous singlet-pair magnetotransport as opposed 
to conventional NS junctions~\cite{MTF} 
where the spin splitting affects single-particle tunneling.     
The predicted bias-energy separation between Andreev enhancement peaks 
for spin-up and spin-down quasiparticles could be implemented for detection 
of ballistic spin-polarized carriers injected from a ferromagnetic source
at low temperatures $T<E_g/k_B$.
We note that the energy and magnetic field behaviour of 
the excess conductance discussed here 
for thick superconductors ($\tilde g\ll 1$) is consistent with 
the experimental findings (see, Refs.~\onlinecite{Giant,Eroms,Eroms1}). 
To observe the predicted spin-dependent effects
the superconducting film must be sufficiently thin. 
Taking $d=5\, nm$ (as in the experiment of Ref.~\onlinecite{MTF}) and 
$k_F\approx 10^6\, cm^{-1}$
typical for high-mobility 2DES 
one finds the requirement for efficient spin splitting to be $\tilde g\approx g>1$.
It can be met in InAs heterostructures where the $g$-factor 
is much larger than 1.  

We thank G. Fagas, V.I. Fal'ko, D. Ryndyk, C. Strunk, and D. Weiss for 
valuable discussions. The work was supported by the 
Deutsche Forschungsgemeinschaft (Forschergruppe 370 "Ferromagnet-Halbleiter-Nanostrukturen").
%%%%%%%%%%%%%%%%%%%%%%%%%%%%%%%%%%%%%%%%%%%%%%%%%%%%%%%%%%%%%%%%%%%%%%%%%%%%%%%%%%%%%%%%%%%%%%%%%

\newpage 

\end{document}